\documentclass[review]{elsarticle}

\usepackage{lipsum}
\makeatletter
\def\ps@pprintTitle{%
 \let\@oddhead\@empty
 \let\@evenhead\@empty
 \def\@oddfoot{}%
 \let\@evenfoot\@oddfoot}

\begin{document}

\begin{frontmatter}

\title{A depleted and numerical model for pulsed Gaussian wave type II second harmonic generation}


\author[mymainaddress]{Mostafa Mohammad Rezaee}

\author[mysecondaryaddress]{Mohammad Sabaeian\corref{mycorrespondingauthor}}
\cortext[mycorrespondingauthor]{Corresponding author}
\ead{sabaeian@scu.ac.ir}

\author[mysecondaryaddress]{Alireza Motazedian}
\author[mysecondaryaddress]{Fatemeh Sedaghat Jalil-Abadi}

\address[mymainaddress]{Data Science program, Bowling Green State University, Bowling Green, OH, US}
\address[mysecondaryaddress]{Department of Physics, Shahid Chamran University of Ahvaz, Ahvaz, Khuzestan, Iran}

\begin{abstract}
This paper presents a three-dimensional (3-D) and spatio-temporal dependent nonlinear wave model to explain the generation of pulsed Gaussian Second Harmonic Waves (SHW). We solve numerically three coupled equations describing the type II Second Harmonic Generation (SHG) in cylindrical KTP crystals under the assumption of fundamental waves depletion, two for ordinary and one for extraordinary Fundamental Waves (FW). The results are attained by a homemade code written in FORTRAN. The results depict the efficiency of the SHG process with the conversion of FW energy to SHW energy, while SHW keeps the same Gaussian profile as FW. Furthermore, the results examine the effect of pulse energy, beam spot size, and therefore interaction length on SHW efficiency.
\end{abstract}

\begin{keyword}
\texttt Second harmonic generation
\sep Pulsed Gaussian wave
\sep KTP
\end{keyword}

\end{frontmatter}



\section{Introduction}
Over the past century, the movement of electromagnetic radiations through the medium has been modeled and examined \cite{gillen2009application_1}. As a result of substantial advancement in computer speed, the accurate and detailed solutions of sophisticated electromagnetic equations have been progressed \cite{guha2005description_2,gillen2006optical_3,hsu1994stratton_4,li2005focal_5}. Modeling of Gaussian beam propagation has been considered as a significant field of research under a wide variety of circumstances \cite{zhou2008analytical_6,duan2005polarization_7}, mainly because laser structures are usually arranged in a particular configuration to generate output beams that are Gaussian in which the laser is said to be operating in the fundamental transverse mode.
\\
In the first years of laser invention by ion argon gas lasers and copper vapor lasers, green light with 532 nm wavelength was generated. Greenlight lasers have a widespread application in the medicine and telecommunication industry, such as green light laser therapy \cite{chernoff2007tissue_8}, underwater communication and ocean exploration \cite{xu2005110_9,wiener1980role_10}, and spectroscopy \cite{lu2000raman_11}. However, because of the expensive setup and maintenance of lasers mentioned above, scientists discovered an invaluable scientific and economical approach called mid-infrared (1–10 micron) frequency upconversion for green light generation using nonlinear solid-state crystals \cite{neely2012broadband_12} such as KTP. In this method, the emitted Gaussian infrared laser beam with wavelength 1064 nm from crystal Nd: YAG is converted into a green light laser beam with wavelength 532 nm using a nonlinear process is called second harmonic generation (SHG). 
\\
The nonlinear crystal KTP (Potassium Titanyl Phosphate) with chemical formulation KTiOPO4 is a brilliant crystal with a high nonlinear conversion coefficient, relatively high thermal conductivity \cite{piskarskas1999noncollinear_13} and high quantum efficiency that is highly efficient for doubling the frequency of the fundamental wave (FW). At mid or low pumping power, the efficiency of the SHG is quite remarkable since it preserves its nonlinear properties and light quality \cite{zhang2008increased_14}. However, at high pumping power, due to the heating of crystal, the refractive indices change with increasing temperature \cite{weise2000continuous_15}. Hence, if the crystal temperature is not monitored, the second harmonic efficiency falls dramatically \cite{zheng2001influence_16}.
\\
Recently a modeling of type II SHG in a single pass cavity under continuous-wave Gaussian beams has been presented \cite{sabaeian2010investigation_17}. In their studies, Sabaeian et al. investigated the equations of the ideal field in a two-pass cavity for continuous Gaussian beam \cite{sabaeian2015temperature_17a}. They coupled the field equations with the equations of heat and phase mismatch and compared the resulting model with the experimental results \cite{sabaeian2014heat_17b}. In this work, the SHG equations with heat and phase equations have been coupled. This model has been examined experimentally to increase the nonlinear conversion efficiency \cite{regelskis2012efficient_18}. By using the sum frequency generation approach \cite{sabaeian2010investigation_17}, a depleted model for pulsed Gaussian wave type II SHG, with ignoring thermal effects and phase mismatching, are solved numerically. \\
In our model, the fundamental beams and type II configuration of the SHG beam are coupled ideally that the thermal effects are ignored, and absorption coefficients are equaled to zero. In this case, the KTP is exposed to the polarized FW with the same frequencies, i.e., one polarized FW is propagated along with ordinary, and another along the extraordinary direction, and the third beam with doubled frequency is generated in the extraordinary direction. Cylindrical coordinates have been used to profit from the azimuthal symmetry of the pumping source profile. Consequently, the temporal differential equations are discretized only on r and z coordinates. 
\\
Our three coupled equations model is adopted to depict pulsed type II configuration. One fundamental beam equation with ordinary polarization, one fundamental beam equation with extraordinary polarization, and one equation for second harmonic wave (SHW) have been considered. Thermal effects such as thermally phase mismatching and thermal lensing have been neglected. Making an analytical solution with great sophisticated and coupled equations is almost highly unlikely. On the other hand, providing a possible and simultaneous solution to these differential equations is RAM storage and a time-consuming procedure. So a numerical approach via the finite difference method has been developed with tremendous creativity and strategy. A homemade code in which the coupled equations can be solved with a personal computer was written in FORTRAN in the Linux Ubuntu operating system.

\section{Theory}
The pulsed SHG equations with a classical formalization could be directly extracted from Maxwell’s equations since the electromagnetic fields can be explained and justified via Maxwell’s equations. At first, the Helmholtz’s equation in steady state for a dispersive and nonlinear medium is considered, which is given as follows \cite{Boyd2003Nonlinear_19}, and that is a monochromatic polarized field:

\begin{equation}
\nabla^{2} \vec{E}_{n}(r, z, t)+\frac{\omega_{n}^{2}}{c^{2}} \varepsilon^{(l)}\left(\omega_{n}\right) \cdot \vec{E}(r, z, t)=-\frac{\omega_{n}^{2}}{\varepsilon_{0} c^{2}} \vec{P}_{n}^{N L}(r, z, t)
\label{eq:1}
\end{equation}

Where $\vec{E}_{n}(r, z, t)$ is an electric field,  $\vec{P}_{n}^{N L}(r, z, t)$ is nonlinear polarization, and $c$ and $\varepsilon^{(l)}$ are light velocity and dielectric tensor, respectively. Regarding absorption at medium, the dielectric tensor is written as:

\begin{equation}
\varepsilon^{l}=\varepsilon_{r}^{l}+i \varepsilon_{i m}^{l}
\label{eq:2}
\end{equation}

The field equation and nonlinear polarization are given as follows \cite{feng2012efficient_20}:

\begin{equation}
P_{n}^{N L}(r, z, t)=P_{n}^{N L}(r, z) e^{i k_{n} z-i \omega_{n} t}+C . C
\label{eq:3}
\end{equation}

\begin{equation}
E_{n}(r, z, t)=E_{n}(r, z) e^{i k_{n} z-i \omega_{n} t}+C . C
\label{eq:4}
\end{equation}

Where $C.C$ is complex conjugate, $\omega$ is the angular frequency, and $k$ is wave number. Because of azimuthal symmetry, the only r and z coordinates are kept, and the Laplacian is taken only in directions above into account, i.e. $\nabla^{2}=\nabla_{r}^{2}+\nabla_{z}^{2}$.
So after some derivation and simplification following expression is reached:

\begin{equation}
\frac{d \vec{E}_{n}(r, z, t)}{d z}-\frac{i}{2 k} \nabla_{r}^{2} \vec{E}_{n}(r, z, t)+\frac{\gamma}{2} \vec{E}_{n}(r, z, t)=\frac{i \omega_{n}}{2 n_{n} \varepsilon_{0} c} \vec{P}_{n}^{N}(r, z, t)
\label{eq:5}
\end{equation}

Where $\varepsilon_{\mathrm{o}}=8.85 \times 10^{-12} C^{2} / \mathrm{Nm}^{2}$ is vacuum permittivity and $\gamma=\varepsilon_{i m}\frac{\omega}{n c}$ is absorption coefficient.

On the temporal derivation of field equations, and the nonlinear sources of fundamental beams and second harmonic beam are introduced as follow \cite{sabaeian2010investigation_17,Boyd2003Nonlinear_19}:

\begin{equation}
P_{1}^{N L}=4 \varepsilon_{0} d_{e f f} E_{3}(r, z) E_{2}^{*}(r, z) e^{i\left(k_{2}-k_{3}\right) z} e^{-i \omega_{1} t}
\label{eq:6}
\end{equation}

For $\omega_1 = \omega$ , and

\begin{equation}
P_{2}^{N L}=4 \varepsilon_{0} d_{e f f} E_{3}(r, z) E_{1}^{*}(r, z) e^{i\left(k_{1}-k_{3}\right) z} e^{-i \omega_{2} t}
\label{eq:7}
\end{equation}

For $\omega_2 = \omega$ , and for $\omega_3 = 2\omega$

\begin{equation}
P_{3}^{N L}=4 \varepsilon_{0} d_{e f f} E_{1}(r, z) E_{2}^{*}(r, z) e^{i\left(k_{1}+k_{2}\right) z} e^{-i \omega_{3} t}
\label{eq:8}
\end{equation}

A set of coupled equations is written as below

\begin{equation}
\frac{n_{1}}{c} \frac{d E_{1}}{d t}+\frac{d E_{1}}{d z}-\frac{i c}{2 n_{1} \omega} \nabla_{r}^{2} E_{1}+\frac{\gamma_{1}}{2} E_{1}=\frac{2 i \omega}{n_{1} c} d_{eff} E_{2}^{*} E_{3} e^{-i \Delta k z}
\label{eq:9}
\end{equation}

\begin{equation}
\frac{n_{2}}{c} \frac{d E_{2}}{d t}+\frac{d E_{2}}{d z}-\frac{i c}{2 n_{2} \omega} \nabla_{r}^{2} E_{2}+\frac{\gamma_{2}}{2} E_{2}=\frac{2 i \omega}{n_{2} c} d_{eff} E_{1}^{*} E_{3} e^{-i \Delta k z}
\label{eq:10}
\end{equation}

\begin{equation}
\frac{n_{3}}{c} \frac{d E_{3}}{d t}+\frac{d E_{3}}{d z}-\frac{i c}{4 n_{3} \omega} \nabla_{r}^{2} E_{3}+\frac{\gamma_{3}}{2} E_{3}=\frac{4 i \omega}{n_{3} c} d_{eff} E_{1} E_{2} e^{i \Delta k z}
\label{eq:11}
\end{equation}

Where $d_{eff}=2 \chi^{(2)}$ is the effective nonlinear coefficient \cite{Boyd2003Nonlinear_19}. In type II SHG with phase matching $\omega_1=\omega_2=\omega$ and $\omega_3=\omega_1+\omega_2=2\omega$ , i.e. the vector mismatching must be equaled to zero $(\Delta k={k_1}+{k_2}-{k_3}=0)$ or phase mismatching should be ignored i.e. $\Delta \phi=\Delta k z=0$. The phase matching can be achieved, if $n^{\omega, o} \omega+n^{\omega, e} \omega=n^{2 \omega, e} 2 \omega$ in above expressions $n_{1}=n^{\omega, o}, n_{2}=n^{\omega, e}, n_{3}=n^{2 \omega, e}$ \cite{sabaeian2010investigation_17}.
\\
All equations to generate a second harmonic wave have been derived; however, dimensionless quantities are used to reduce calculations errors. Therefore, the variables change that is the ratio of generated wave intensity, and initial wave intensity have been introduced which are:

\begin{eqnarray}
\psi_{1}=\frac{E_{1}}{\sqrt{P_{1} / 2 n_{1} c \varepsilon_{0} \pi \omega_{f}^{2}}} \Rightarrow E_{1}=\sqrt{\frac{P_{1}}{2 n_{1} c \varepsilon_{0} \pi \omega_{f}^{2}}} \psi_{1} 
\\
\nonumber
\Rightarrow \eta_{1}= \left|\psi_{1}\right|^{2}=\frac{2 n_{1} c \varepsilon_{0}\left|E_{1}\right|^{2}}{P_{1} / \pi \omega_{f}^{2}}=\frac{I_{1}}{I_{1}(0)}
\label{eq:12}
\end{eqnarray}

\begin{eqnarray}
\psi_{2}=\frac{E_{2}}{\sqrt{P_{2} / 2 n_{2} c \varepsilon_{0} \pi \omega_{f}^{2}}} \Rightarrow E_{2}=\sqrt{\frac{P_{2}}{2 n_{2} c \varepsilon_{0} \pi \omega_{f}^{2}}} \psi_{2}
\\
\nonumber
\Rightarrow \eta_{2}=\left|\psi_{2}\right|^{2}=\frac{2 n_{2} c \varepsilon_{0}\left|E_{2}\right|^{2}}{P_{2} / \pi \omega_{f}^{2}}=\frac{I_{2}}{I_{2}(0)}
\label{equ:13}
\end{eqnarray}

\begin{eqnarray}
\psi_{3}=\frac{E_{3}}{\sqrt{P_{3} / 2 n_{3} c \varepsilon_{0} \pi \omega_{f}^{2}}} \Rightarrow E_{3}=\sqrt{\frac{P_{3}}{2 n_{3} c \varepsilon_{0} \pi \omega_{f}^{2}}} \psi_{3}
\\
\nonumber
\Rightarrow \eta_{3}=\left|\psi_{3}\right|^{2}=\frac{2 n_{3} c \varepsilon_{0}\left|E_{3}\right|^{2}}{P_{3} / \pi \omega_{f}^{2}}=\frac{I_{3}}{I_{1}(0)+I_{2}(0)}
\label{eq:14}
\end{eqnarray}

In which the quantity $P_i$ and $\eta_{i}$ with $i =$ 1, 2, and 3, gives the power and intensity efficiency for existence waves, respectively, and $\omega_f$ denotes the fundamental beam spot size. As the fundamental waves have the same frequency, their power with orthogonal polarization is equal, as well. Hence, in the $z=0$ plane, the power of the fundamental wave $P_1 = P_2 = P$ and via the SHG approach the power of final wave equal to $P_3 = 2P$ . Replacing the change of variable Eq. (11) to Eq. (13) denoted above, and after some mathematical treatments, the three type II SHG equations are achieved as below:

\begin{equation}
\frac{n_{1}}{c} \frac{d \psi_{1}}{d t}+\frac{d \psi_{1}}{d z}-\frac{i c}{2 n_{1} \omega} \frac{1}{r} \frac{d \psi_{1}}{d r}-\frac{i c}{2 n_{1} \omega} \frac{d^{2} \psi_{1}}{d r^{2}}+\frac{\gamma_{1}}{2} \psi_{1}=\frac{i}{L} \psi_{2}^{*} \psi_{3} e^{-i \Delta \phi}
\label{eq:15}
\end{equation}

\begin{equation}
\frac{n_{2}}{c} \frac{d \psi_{2}}{d t}+\frac{d \psi_{2}}{d z}-\frac{i c}{2 n_{2} \omega} \frac{1}{r} \frac{d \psi_{2}}{d r}-\frac{i c}{2 n_{2} \omega} \frac{d^{2} \psi_{2}}{d r^{2}}+\frac{\gamma_{2}}{2} \psi_{2}=\frac{i}{L} \psi_{1}^{*} \psi_{3} e^{-i \Delta}
\label{eq:16}
\end{equation}

\begin{equation}
\frac{n_{3}}{c} \frac{d \psi_{3}}{d t}+\frac{d \psi_{3}}{d z}-\frac{i c}{4 n_{3} \omega} \frac{1}{r} \frac{d \psi_{3}}{d r}-\frac{i c}{4 n_{3} \omega} \frac{d^{2} \psi_{3}}{d r^{2}}+\frac{\gamma_{3}}{2} \psi_{3}=\frac{i}{L} \psi_{1} \psi_{2} e^{i \Delta \phi}
\label{eq:17}
\end{equation}

The interaction length L is defined by

\begin{equation}
L=\left(\frac{n_{1} n_{2} n_{3} c^{3} \varepsilon_{0} \pi \omega_{f}^{2}}{4 P \omega^{2} d_{e f f}^{2}}\right)^{\frac{1}{2}}
\label{eq:18}
\end{equation}

As far as our knowledge is concerned, there has not been any specific research on this quantity yet. That is a particular length in which the most nonlinear interaction fulfills. In other words, this quantity clearly presents a profound perception of effective parameters within nonlinear leasers. For instance, nonlinear effective coefficient $d_{eff}$ and refractive indices are the characteristics of nonlinear crystal i.e. with changing these parameters, the efficiency of SHG can be foreseen approximately in advance.
\\
As it was mentioned previously, the Gaussian beams are of interest in this study. Thus, we presume a Gaussian beam for the laser source or fundamental beams as boundary conditions at the crystal input plane in which $\psi_{1}(r, z=0)=\psi_{2}(r, z=0)=\exp \left(-r^{2} / \omega_{f}^{2}\right)$ and $\psi_{3}(r, z=0)=0$.

Time-dependent boundary conditions at $z=0$ for fundamental and second harmonic beams should be taken into account as

\begin{equation}
\psi_{1}(t, r, z=0)=\exp \left[-\left(t / t_{p}\right)^{2}\right] \times \exp \left(-r^{2} / \omega_{f}^{2}\right)
\label{eq:19}
\end{equation}

\begin{equation}
\psi_{2}(t, r, z=0)=\exp \left[-\left(t / t_{p}\right)^{2}\right] \times \exp \left(-r^{2} / \omega_{f}^{2}\right)
\label{eq:20}
\end{equation}

\begin{equation}
\psi_{3}(t, r, z=0)=0
\label{eq:21}
\end{equation}

Such that $\psi_{1}(t, r, z=0)=\psi_{2}(t, r, z=0) \approx 0$ and $t_p$ is the duration time of pulses.

\section{Results and discussion}
The fields coupled equations Eq. (14)-(16) are solved ideally by a numerical approach and with a homemade code written in FORTRAN and run by Linux Ubuntu operating system. The fields coupled equations are discretized at cylindrical coordinates using the finite difference method (FDM). The discretization has been done for temporal parts of Eq. (14)-(16) using backward FDM, for spatial parts along the crystal axis forward FDM, and for spatial parts along radial direction central FDM. The optical parameters of the crystal and the other physical constant have been listed in Table 1 and Table 2, respectively.

\begin{table}[ht]
\centering \caption{The optical crystal parameters of KTP}
\label{tab:1}\vskip .1in
\begin{tabular}{|c|c|c|}\hline
Length                          & L=2 cm                                        & \cite{seidel1997numerical_21} \\ \hline
Radius                          & r=2 mm                                        & \cite{seidel1997numerical_21} \\ \hline
Effective nonlinear coefficient & $d_{eff}=7.3$ pmV$^{-1}$                      & \cite{sabaeian2010investigation_17} \\ \hline
Ordinary refractive index       & $n^{o, \omega}=1.8296$                        & \cite{kato1991parametric_22} \\ \hline
Extraordinary refractive index  & $n^{\mathrm{e}, \omega}=1.7466$               & \cite{kato1991parametric_22} \\ \hline
Extraordinary refractive index  & $n^{o, 2 \omega}=1.7881$                      & \cite{kato1991parametric_22} \\ \hline
Crystal cutting angles          & $\theta=90^{\circ},\; \varphi=24.77^{\circ}$  & \cite{sabaeian2010investigation_17} \\ \hline
\end{tabular}
\end{table}

\begin{table}[ht]
\centering \caption{The physical constant used at coupled field equations}
\label{tab:2}\vskip .1in
\begin{tabular}{|c|c|}\hline
Fundamental wavelength                     & $\lambda_{1}=1064 \;nm$     \\ \hline
Second harmonic wavelength                 & $\lambda_{2}=532 \;nm$      \\ \hline
Pulse duration                             & $t_{p}=50 \;\mu s$          \\ \hline
Pulse repetition frequency                 & f = 4000 Hz                 \\ \hline
Pulse energy                               & E = 0.45 J                  \\ \hline
Beam spot size                             & $\omega_{f}=80 \;\mu m$     \\ \hline
Number of time steps                       & Nt = 2511                   \\ \hline
Number of steps in radial direction        & Nr = 120                    \\ \hline
Number of steps in longitudinal direction  & Nz = 12000                  \\ \hline
\end{tabular}
\end{table}

\begin{figure}[!htbp]
\centering
\includegraphics[width = 4 in ]{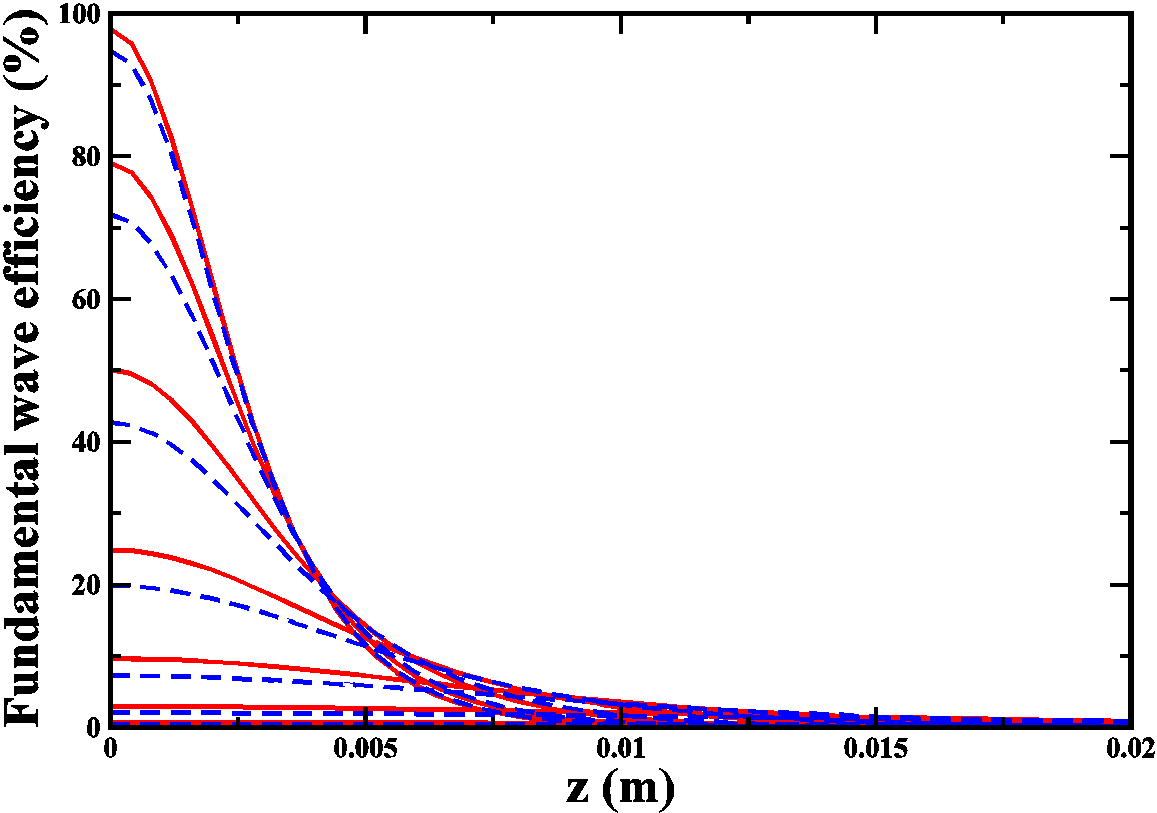}
\caption{Efficiency of the fundamental wave along the crystal axis for several times: from the top to the bottom solid curves are $t= 25ms, 75ms, 100ms$ ,and from the top to the bottom dashed curves are $t=125ms, 150ms, 175ms, 200ms$.}
\label{fig:1}
\end{figure}

Figure \ref{fig:1} shows the efficiency of FW along the crystal axis for different times from zero to $200ms$. The solid (red) curves are for times between $25ms$ and $100ms$ starting from zero percent at $t=25ms$ to 100 percent at $t=100ms$, and dashed (blue) curves are related to the times from $125ms$ to 200ms beginning from almost 90 percent reaching zero at $t=200ms$.

\begin{figure}[!htbp]
\centering
\includegraphics[width = 4 in ]{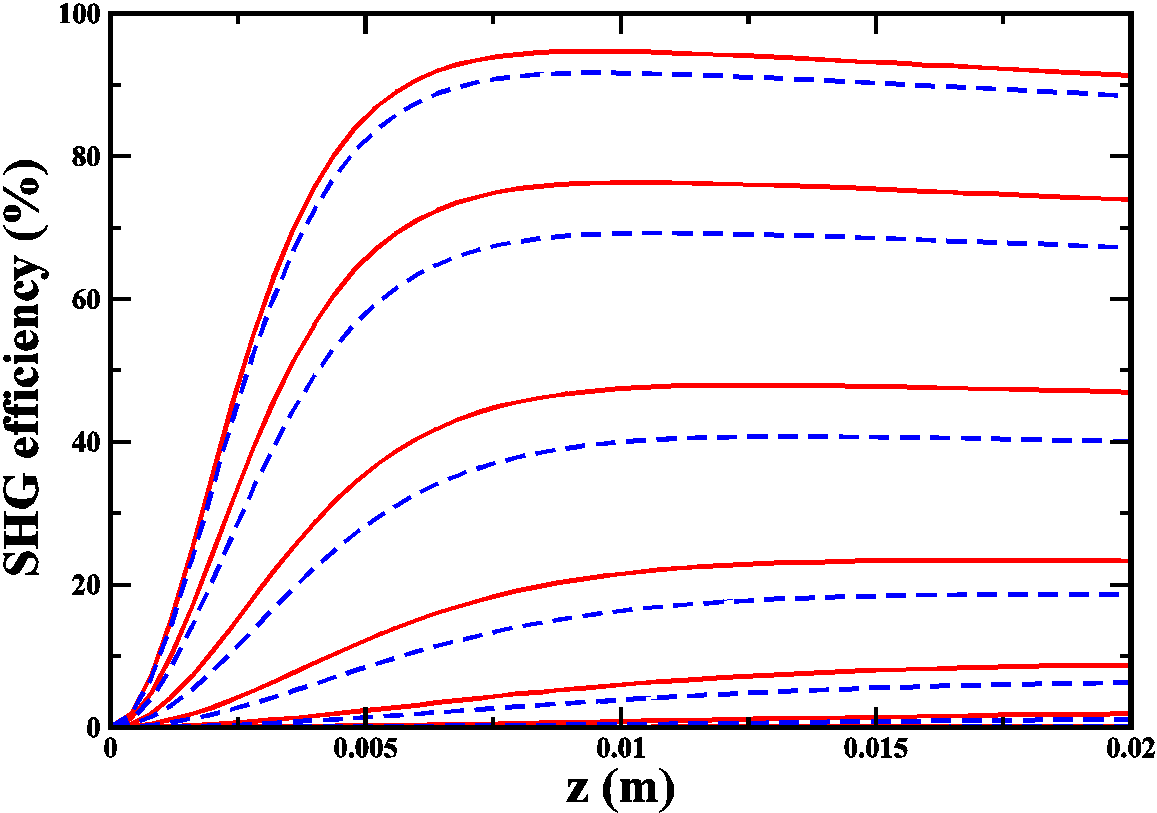}
\caption{Efficiency of the second harmonic generation along the crystal length at several times as used in Figure \ref{fig:1}}
\label{fig:2}
\end{figure}

However, for every time duration, the FW efficiency drops to zero steadily, and it can be inferred that the FW interchange its energy to the SHW irreversibly, which can be perceived and understood in Figure \ref{fig:2}. In this case, the efficiencies of SHW have been depicted via time durations in Figure \ref{fig:1} the solid (red) curves reveal the increase in the second harmonic field, and the dashed (blue) curves display the decrease in it. Corresponding to this figure, the efficiency of SHW reaches its greatest amount, which is 100 percent at $t=2t_p$. After moving almost $z=5mm$ through the crystal, the efficiency falls slightly due to the absorption of the SHW in the crystal.

\begin{figure}[!htbp]
\centering
\includegraphics[width = 4 in ]{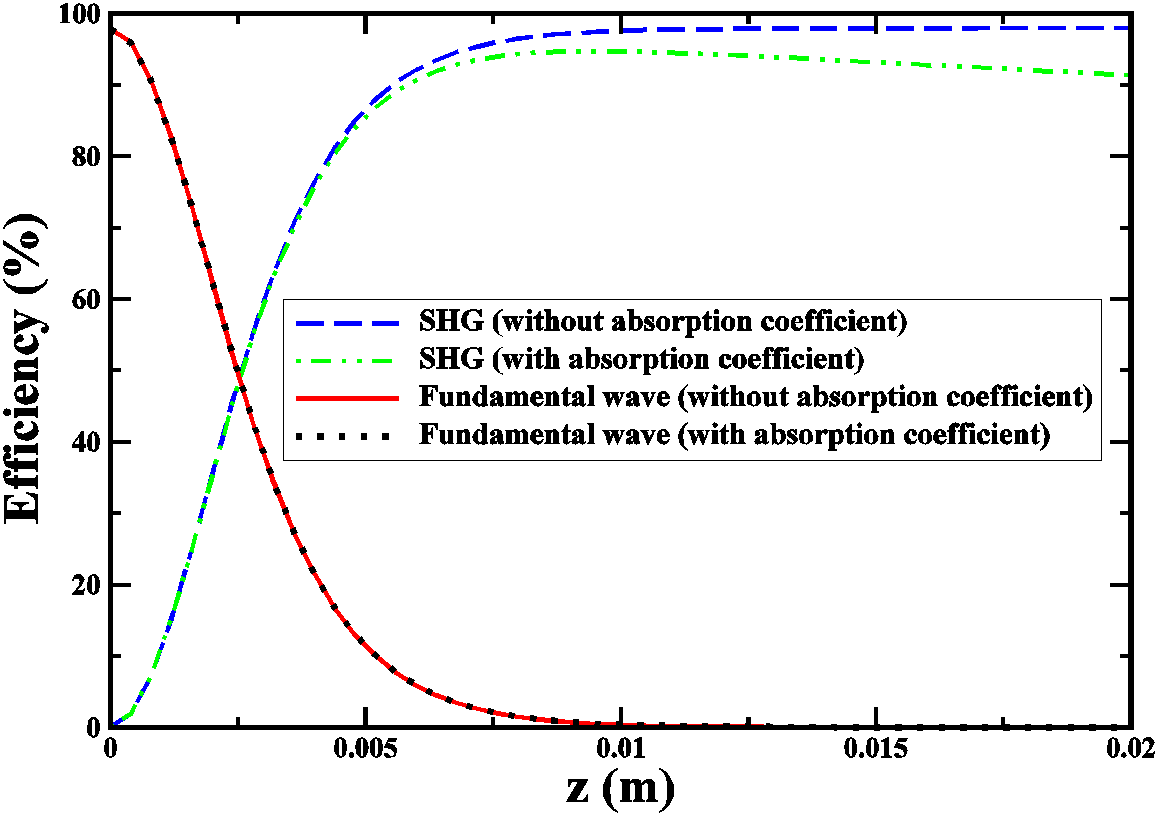}
\caption{Efficiency of the second harmonic generation and the fundamental wave along the crystal length.}
\label{fig:3}
\end{figure}

\begin{figure}[!htbp]
\centering
\includegraphics[width = 4 in ]{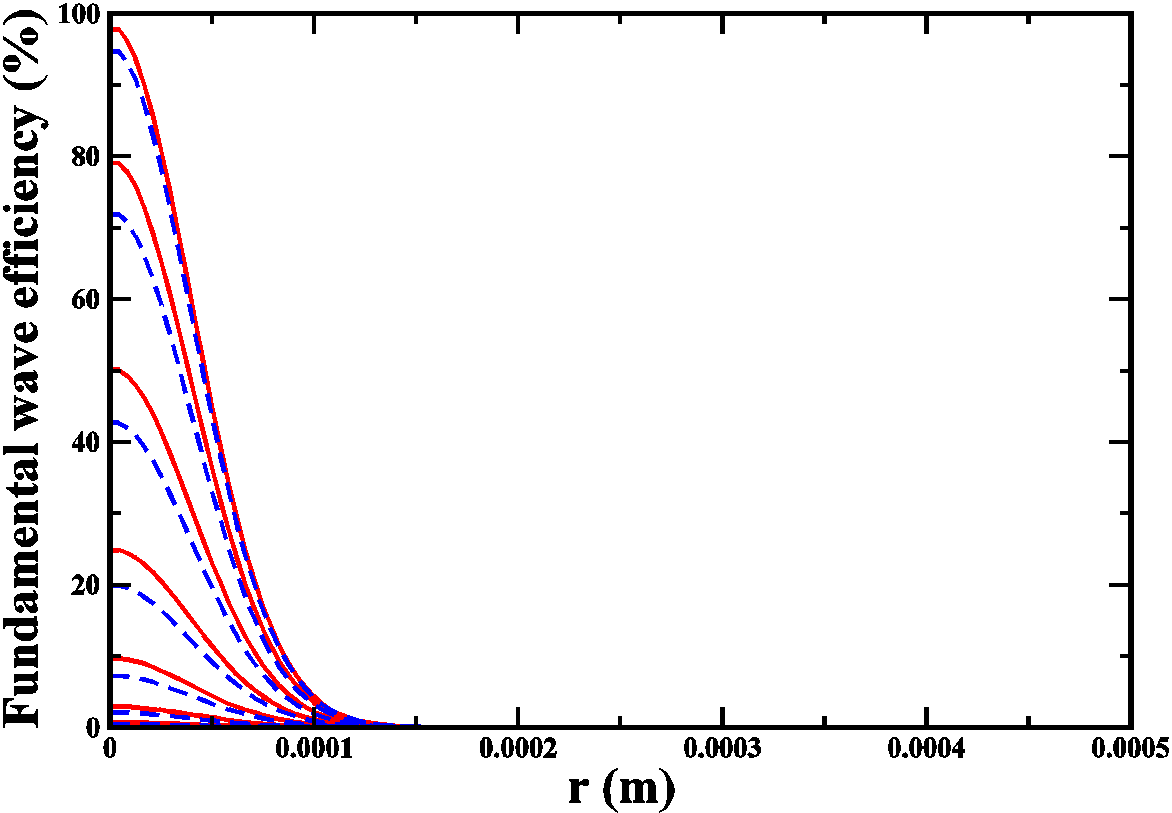}
\caption{Efficiency of the fundamental wave in the radial direction for input surface of the crystal. Various curve corresponds to times mentioned as used in Figure \ref{fig:1}.}
\label{fig:4}
\end{figure}

Figure \ref{fig:3} displays the variations of the FW and SHW along the crystal axis at $t=2t_p$ in which the pulse reaches its maximum energy. Two different cases are compared, ignoring the optical absorption waves shown by the dotted curve for the FW and the dashed-dotted curve for the SHW efficiencies. On the other hand, the optical absorption waves are revealed by the solid curve for the FW and the dashed curve for the SHW efficiencies. As explained here, the conversion energy between FW and SHW happens at a distance of 5mm compared to the crystal length. Consequently, for the crystal and the beams used in this article, the depleted formalism is necessarily essential, and thus constant beam approximation for FW cannot be correct anymore.
\\
Figure \ref{fig:4} shows the profile of the efficiency of the FW radially at the entrance surface of the crystal (z = 0). The incident FW onto the crystal face has a Gaussian profile. Before starting the energy interchange, the incident waves’ efficiency entrance face of the crystal is 100 percent, where falls gradually towards the crystal lateral surface.

\begin{figure}[!htbp]
\centering
\includegraphics[width = 4 in ]{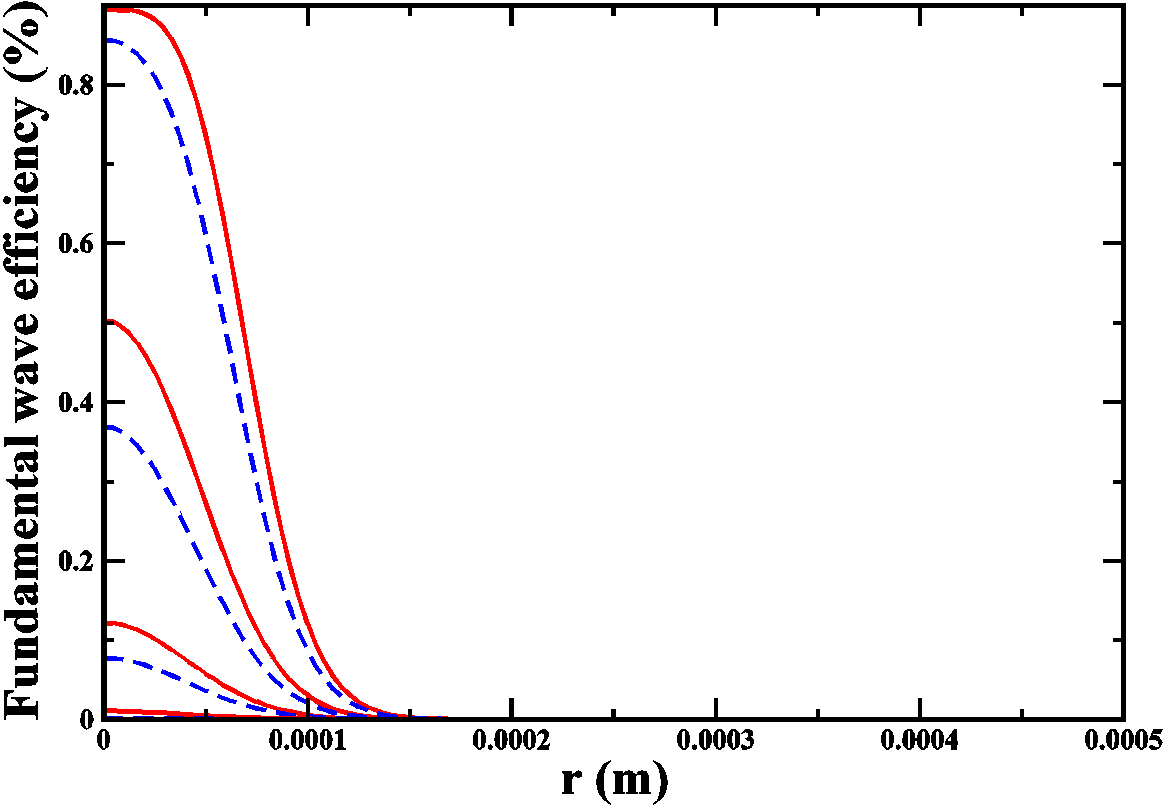}
\caption{Efficiency of the fundamental wave in the radial direction for output surface of the crystal.}
\label{fig:5}
\end{figure}

At the exit face, Figure \ref{fig:5} illustrates the FW efficiency is not to be considered and reaches 0.5 percent. On the other hand, the Gaussian SHW is generated in the crystal due to energy conversion.

\begin{figure}[!htbp]
\centering
\includegraphics[width = 4 in ]{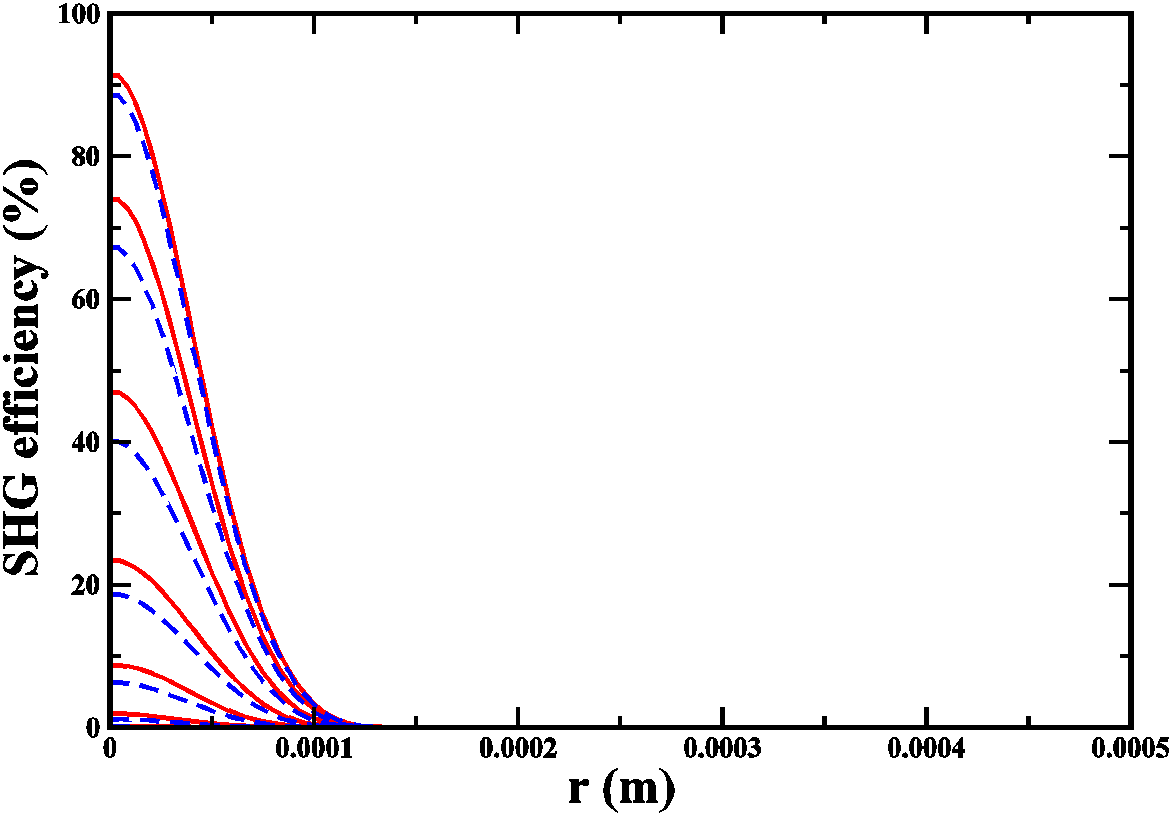}
\caption{Efficiency of the SHG in the radial direction for output surface of the crystal. }
\label{fig:6}
\end{figure}

Figure \ref{fig:6} demonstrates the SHW transverse Gaussian profile at the output surface of crystal at z = 2 $cm$ , as it is predicted for an ideal and depleted mechanism, the SHW efficiency at the end face of the crystal is equal to 100\%, approximately, indicating all the FW energy are converted to the SHW energy, regarding this fact that the SHW has the same transverse Gaussian profile.

\begin{figure}[!htbp]
\centering
\includegraphics[width = 4 in ]{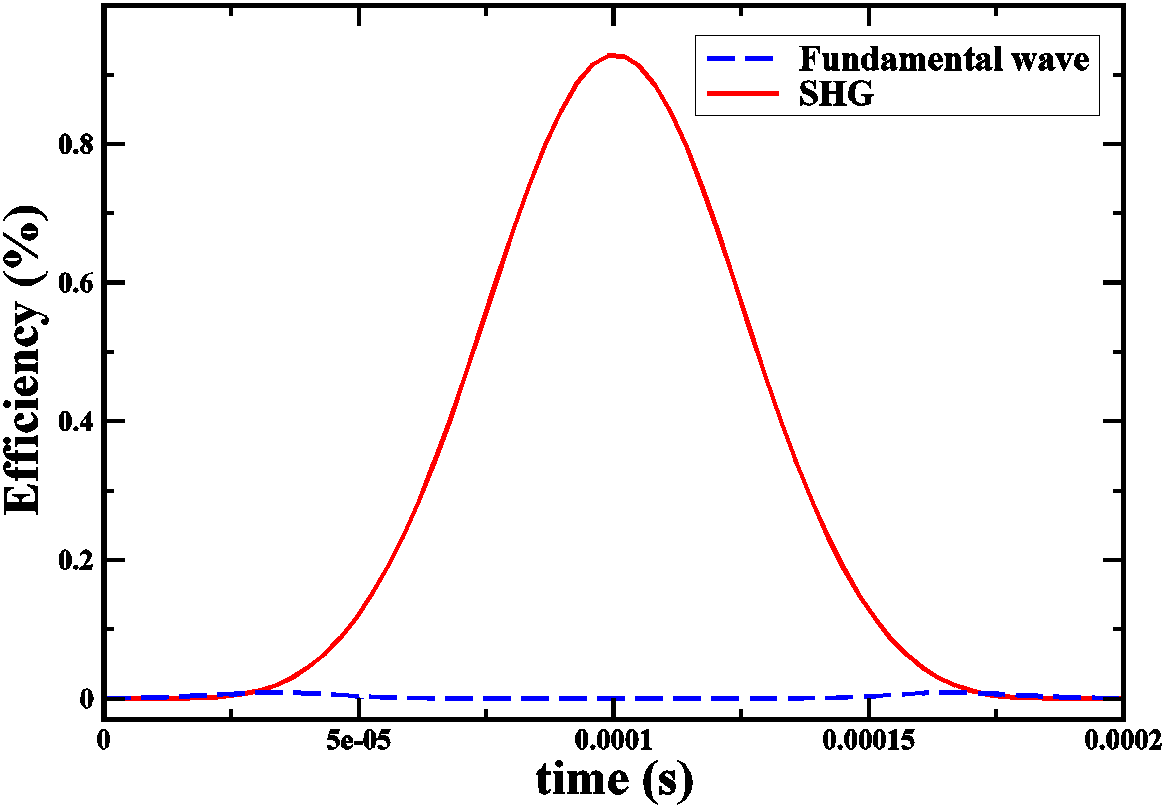}
\caption{The efficiency of the fundamental wave (dashed curve) and second harmonic generation (solid curve) at the output face of the crystal from t=0 to t=4$t_p$.}
\label{fig:7}
\end{figure}

\begin{figure}[!htbp]
\centering
\includegraphics[width = 4 in ]{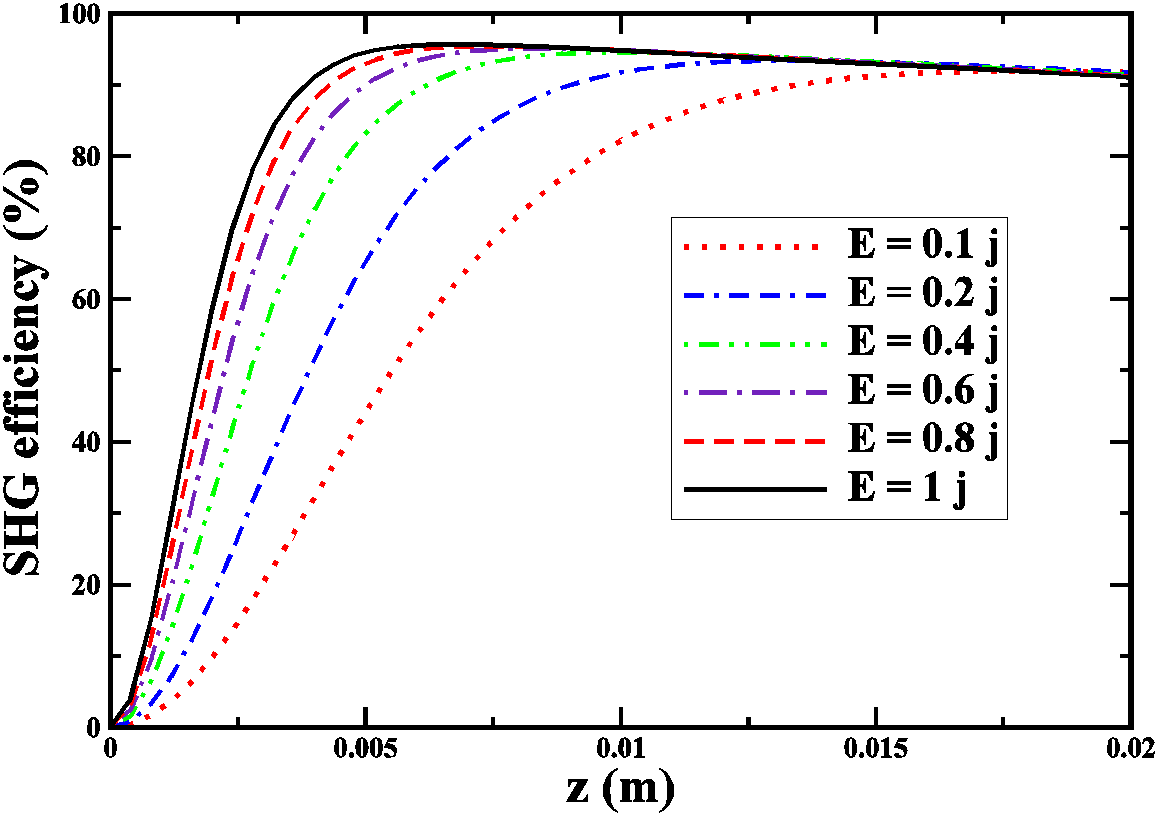}
\caption{Efficiency of the second harmonic generation along the crystal length for different energies.}
\label{fig:8}
\end{figure}

Figure \ref{fig:7} represents the temporal variation of the FW and SHW efficiencies at the central point of the output surface r = 0 of crystal. It is emphasized that the whole of the FW energy (dashed curve) is converted to the SHW energy (solid curve) in an ideal state.
\\
The variations of SHW efficiency is investigated by altering the energy of pulses, in which the $0.1 j,\;0.2 j,\;0.4 j,\;0.6 j,\;0.8 j,$ and $1j$ are used, and the resultant data is exhibited in Figure \ref{fig:8}. As the Figure \ref{fig:8} shows and inferred via Eq. (17), the higher energy causes the smaller interaction length and thus the quicker energy conversion. To be more precise, for higher energies, the conversion of energy between FW and SHW accrues at closer distances to the input surface.

\section{Conclusion}
In this work, the SHG type II configuration has been applied to generate SHW with the Gaussian profile like its creating wave, in this case, the three coupled equation which was two ordinary and extraordinary FWs and one extraordinary SHW, are interacting in the crystal, were solved numerically at the same time. On the other hand, to describe the SHG process, a three-dimensional (3-D) and spatio-temporal dependent nonlinear waves model assuming the depletion of fundamental waves was considered. The results were attained by a homemade code written in FORTRAN that was run in Linux Ubuntu operating system. The optical absorption for both FW and SHW has been considered to obtain more reliable results and closer to physical reality. The results depicted how energy was converted during the waves’ interaction from FW to SHW and how much energy was converted along the crystal axis.

\bibliography{mybibfile}

\end{document}